\documentclass[aps,prb,twocolumn]{revtex4-1}

\usepackage{graphicx}
\usepackage{multirow}
\usepackage{hyperref}
\usepackage{amsmath}
\usepackage{color}
\usepackage{ulem}
\usepackage{setspace}

\begin{document}

\title{Energy derivatives in real--space diffusion Monte Carlo}

\author{Jesse van Rhijn\,}
\email{j.vanrhijn@utwente.nl}
\affiliation{MESA+ Institute for Nanotechnology, University of Twente, P.O.  Box 217, 7500 AE Enschede, The Netherlands}
\author{Claudia Filippi\,}
\email{c.filippi@utwente.nl}
\affiliation{MESA+ Institute for Nanotechnology, University of Twente, P.O.  Box 217, 7500 AE Enschede, The Netherlands}
\author{Stefania De Palo\,}
\email{depalo@iom.cnr.it}
\affiliation{CNR--IOM DEMOCRITOS, Istituto Officina dei Materiali, and SISSA Scuola Internazionale Superiore di Studi Avanzati, Via Bonomea 265, I--34136 Trieste, Italy}
\author{Saverio Moroni\,}
\email{moroni@iom.cnr.it}
\affiliation{CNR--IOM DEMOCRITOS, Istituto Officina dei Materiali, and SISSA Scuola Internazionale Superiore di Studi Avanzati, Via Bonomea 265, I--34136 Trieste, Italy}

\begin{abstract}
We present unbiased, finite--variance estimators of energy derivatives 
for real--space diffusion Monte Carlo calculations within the fixed--node 
approximation. The derivative $d_\lambda E$ is fully consistent with 
the dependence $E(\lambda)$ of
the energy computed with the same time step. We address 
the issue of the divergent variance of derivatives related to variations 
of the nodes of the wave function, both by using a regularization for 
wave function parameter gradients recently proposed in variational Monte 
Carlo, and by introducing a regularization based on a coordinate 
transformation. The essence of the divergent variance problem is 
distilled into a particle-in-a-box toy model, where we demonstrate 
the algorithm.

\end{abstract}

\maketitle

\section{Introduction}

Variational Monte Carlo (VMC) and diffusion Monte Carlo (DMC)
are numerical stochastic approaches based on a real--space representation of a
correlated trial wave function to study many--body quantum systems,
including electronic structure problems.
VMC calculates expectation values of quantum operators on the
trial function, which in turn is optimized via minimization of 
a suitable cost function such as the variational energy.
DMC further improves the VMC results through a stochastic implementation of 
the power method, which projects the lowest--energy component of the
trial function. Its accuracy, in the fixed--node (FN) approximation almost 
invariably adopted to avoid the sign problem, is ultimately limited 
by the error in the nodal surface of the trial function.\,\cite{reviews} 

In the last decade, the efficient calculation of analytic energy
derivatives,\,\cite{ad,assaraf} leveraging modern optimization 
methods,\,\cite{opt,linear,sr} spawned impressive progress in 
both accuracy and scope of the VMC 
method.\,\cite{qmcpack,turborvb,casino,champ}

DMC largely benefits from advances in VMC because improved trial 
functions tend to have better nodes. However, it would be desirable
to have efficient and unbiased estimators of derivatives in DMC as well,
to perform such tasks as the direct optimization of the 
nodal surface or DMC structural relaxation.
This is still an open issue, with
the latest developments featuring uncontrolled approximations 
and/or very low efficiency\,\cite{badinski,corr_walk,vd}.
Here we present an algorithm to calculate unbiased energy derivatives
in FN--DMC with finite variance. 

\section{Energy derivatives}
In both VMC and DMC, the energy is calculated  as
\begin{equation}
E=\int P(R)E_L(R)dR\bigg/\int P(R)dR\equiv\Big\langle E_L\Big\rangle_P 
\label{eq:e}
\end{equation}
where $R$ represents the coordinates of all the particles, 
$E_L(R)=H\Psi(R)\big/\Psi(R)$ is the local energy of
the trial function $\Psi(R)$, and $P(R)$ is proportional to
the underlying probability distribution: in VMC, $P(R)=\Psi^2(R)$,
and, in DMC, $P(R)=\Psi(R)\Phi(R)$ with $\Phi(R)$ the FN solution.
The derivative with respect to a parameter $\lambda$ is
\begin{equation}
d_\lambda E=\Big\langle d_\lambda E_L + (E_L-E)d_\lambda\ln P \Big\rangle_P 
\label{eq:de_bare}
\end{equation}
The variance of this na\"ive estimator is zero if both $\Psi$ and its 
derivative $d_\lambda\Psi$ are 
exact; however, for an approximate trial function, $E_L(R)$ diverges 
at the nodes as $1\big/d(R)$, where $d=|\Psi|\big/\left\|\nabla\Psi\right\|$,
and the variance diverges as well.\,\cite{as,pw}

\subsection{Regularized estimators}
In VMC, this problem was fully solved by Attaccalite and Sorella\,\cite{as} 
with a reweighting scheme, hereafter dubbed AS, whereby one samples the 
square of a modified trial function ${\widetilde \Psi}$ which 
differs from $\Psi$ only for $d$ smaller than a cutoff parameter 
$\epsilon$ and stays finite on the nodal surface of $\Psi$. 
A similar sampling scheme was proposed by Trail\,\cite{trail1}.
The AS estimator has the same average of the bare estimator of 
eq~\ref{eq:de_bare}\, for any value of $\epsilon$, and finite variance. 

An alternative regularized 
estimator, recently proposed by Pathak and Wagner,\,\cite{pw} simply consists in
multiplying the term in brackets of eq~\ref{eq:de_bare}\, by
the polynomial $f_\epsilon(x)=7x^6-15x^4+9x^2$, with $x=d\big/\epsilon$, 
whenever $x<1$. 
This estimator, hereafter dubbed PW, has finite variance
for any finite $\epsilon$ and a bias which vanishes as $\epsilon\to 0$.
The polynomial is chosen in such a way to remove from the bias the linear term,
shown\,\cite{pw} to be $\propto\int_0^1(f_\epsilon-1)dx$, and
to be continuously differentiable in $(0,1)$. The odd parity of fermionic wave 
functions near a node then implies a cubic leading term in the bias. 
Since various values of $\epsilon$ can be used
in the same simulation, the bias can be eliminated at no cost by extrapolation.
The PW estimator has been proposed for parameter gradients in the VMC 
optimization of $\Psi$, but it is equally applicable to VMC interatomic 
forces and, as we will show, to generic derivatives in DMC.

We introduce a third regularized
estimator, that we denote ``warp'' by analogy with the space--warp 
transformation of ref.~\citenum{filippi_umrigar}\, devised to reduce the 
statistical noise of the forces. There, as a nucleus is displaced, a 
transformation is applied to the coordinates of nearby electrons, in such 
a way to maintain the electron--nucleus distances approximately constant.
Here, the goal is to maintain constant the value of $d(R)$ 
when the nodal surface is displaced by a variation of the 
parameter $\lambda$. In this way, the diverging term in 
the local energy does not change, and the variance 
of the derivative is finite. 

\begin{figure}[h!]
\includegraphics[width=5.5cm,angle=-90]{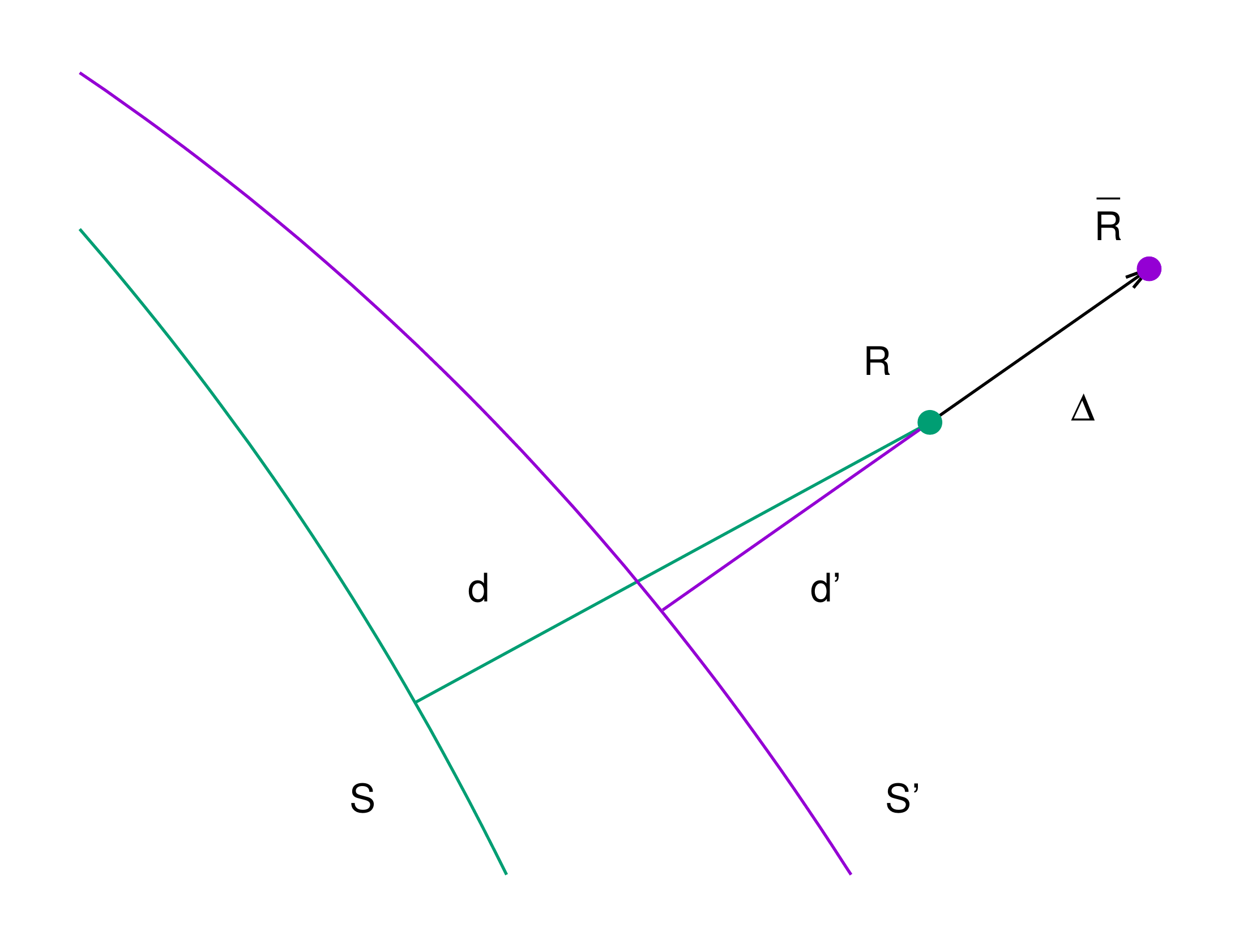}
\caption{
Schematic picture of the warp transformation.
The curve $S$ is the nodal surface for the value $\lambda_0$ of the
parameter $\lambda$. The value of $d$ for the current configuration $R$
is pictorially represented as the distance from $S$
(this is strictly true only if $\Psi$ is linear, 
i.e. close enough to the node).
When a variation of $\lambda$ from $\lambda_0$ to $\lambda'$ shifts $S$ to $S'$, 
the value of $d$ changes to $d'$.
Hence, we displace $R$ by an amount $\Delta=d-d'$ to $\overline{R}$ 
in the direction of ${\rm sign}(\Psi')\nabla\Psi'(R)$, so that the
value of $|\Psi'(\overline{R})|\big/\left\|\nabla\Psi'(\overline{R})\right\|$ 
is approximately equal to $d$.
}
\label{fig:warp}
\end{figure} 
The warp transformation, illustrated in Figure~\ref{fig:warp}\,, is
defined as
\begin{equation}
\overline{R}=R+\left[d(R)-d'(R)\right]{\rm sign}\left(\Psi'(R)\right)
n'(R)u(d(R)) 
\label{eq:warp}
\end{equation}
where primed quantities are calculated for the value $\lambda'$ of the parameter,
$n'$ is the unit vector in the direction of $\nabla\Psi'(R)$, and $u(d)$ is
a cutoff function with support $[0,\epsilon]$ which decreases smoothly 
from 1 to 0, restricting the warp transformation to a region close 
to the nodal surface. We use the quintic polynomial with zero 
first and second derivatives at the boundaries of the support.

For a finite increment of $\lambda$, the energy is
\begin{equation}
E'=\int E'_L(\overline{R}) P'(\overline{R})J dR 
        \bigg/
        \int P'(\overline{R})J
        dR,
\end{equation}
where $J=\det J_{ij}=\det \partial \overline{R}_i\big/\partial R_j$ 
is the Jacobian
of the transformation. The analytic derivative $d_\lambda E$, calculated at
the value of the parameter $\lambda=\lambda_0$, is:
\begin{eqnarray}
&d_\lambda E\big|_{\lambda _0}
=\Big\langle
   \partial_\lambda E_L
  +\nabla E_L
   \cdot \partial_\lambda\overline{R}&\nonumber\\
  &+(E_L-E)
   \left[
     \partial_\lambda\ln(PJ)
    +\nabla\ln P
     \cdot \partial_\lambda\overline{R}
   \right]
 \Big\rangle_P& 
\label{eq:de_warp}
\end{eqnarray}
This warp regularized estimator has finite variance and no bias for any
value of $\epsilon$.

Note that all the functions in 
eq~\ref{eq:de_warp}\, are evaluated at $\lambda=\lambda_0$, where $\overline{R}=R$
and $J=1$. Therefore the warp transformation only contributes to 
the estimator, through the derivatives 
$\partial_\lambda\overline{R}_i\big|_{\lambda_0}$ and $\partial_\lambda \ln J\big|_{\lambda_0}$,
while the sampling is done over one and the same distribution 
$P(R)$ for any parameter we may vary. 

Furthermore, for $\lambda=\lambda_0$, the cofactors of the Jacobi matrix $J_{ij}$
are $M_{ij}=\delta_{ij}$, and the seemingly awkward derivative of the 
Jacobian greatly simplifies,
$\partial_\lambda \ln J\big|_{\lambda_0}=\sum_{ij}M_{ij}\partial_\lambda J_{ij}\big|_{\lambda_0}
=\sum_i\partial_\lambda J_{ii}\big|_{\lambda_0}$,
so that the implementation of eq~\ref{eq:de_warp}\, is not overly complicated.
In particular, most of the derivatives needed are already present --or 
very similar to those already present-- in VMC codes with analytic 
derivatives for structural and full variational optimization. 
The only exceptions are the off--diagonal components of the the Hessian 
$\partial^2\Psi/\partial R_i\partial R_j$ and their derivatives
with respect to $\lambda$, which contribute to $\partial_\lambda J$. We will 
show (heuristically) that the bias incurred by neglecting those terms 
can be extrapolated out at no cost.

\subsection{Variational Monte Carlo}
Before addressing the derivatives in DMC, we compare the three
regularized estimators PW, AS, and warp in VMC. To this purpose, it is
expedient to consider a system stripped of all complexities of
external and interparticle potentials, so that we can focus 
exclusively on the divergence of the local energy at the nodal surface.
Our toy model is a free particle in an elliptic box with hard walls,
meant to represent the configuration of a generic system within a nodal pocket.
Atomic units are used throughout. We choose 
$\Psi=\Psi_0(x,y)=a^2-x^2\big/C-y^2\big/(C-1)$ with $C=[\cosh(1)]^2=2.3810978$, 
which is positive inside the ellipse, vanishes at the border, and is 
not the true ground state. Therefore, the ellipse is defined through the wave
function, and we take the derivative with respect to the parameter $a$, which
changes the 
size of the ellipse at constant eccentricity.
\begin{figure}[h!]
\includegraphics[width=8cm]{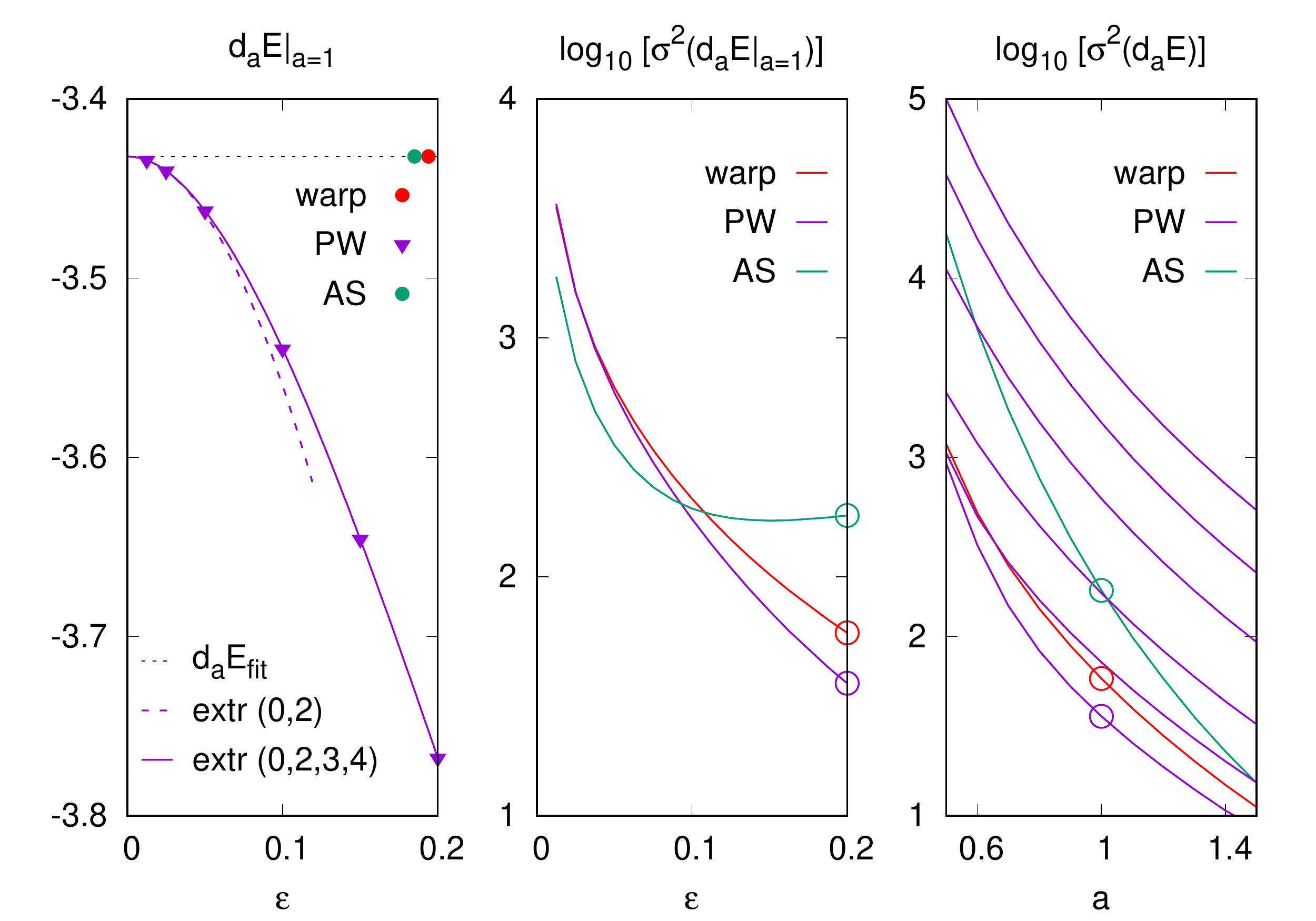}
\caption{VMC calculations of $d_a E$, with integrals performed by 
quadrature. Left panel: the derivative at $a=1$ obtained with
the warp, PW, and AS estimators as a function of the cutoff $\epsilon$, 
compared to the ``exact'' result $d_a E_{\rm fit}$, defined as the derivative 
of a fit to
$E(a)$ calculated separately for several values of $a$. AS and warp, shown
here only near $\epsilon=0.2$, are unbiased; PW can be extrapolated to the
unbiased result either including only the leading term in $\epsilon$,
here $\epsilon^2$, 
on a sufficiently small range (dashed line) or using a sufficiently 
large number of terms on an extended range (solid line). Middle panel: 
variance (in logarithmic scale) of the various regularized estimators as 
a function of the cutoff. As $\epsilon$ vanishes, all schemes regress to 
the infinite variance of the bare estimator. Right panel: variance (in 
logarithmic scale) of the various regularized estimators as a function of 
the parameter $a$. For the warp and AS estimators, $\epsilon=0.2$; 
for PW, data are reported for $\epsilon=0.0125$, 0.025, 0.05, 0.15 
and 0.2 in order of decreasing variance. 
Common data in the middle and the right panels are circled.}
\label{fig:vmc}
\end{figure}

The average and variance of the various estimators, calculated by quadrature,
are shown in Figure~\ref{fig:vmc}\,. None of the regularized estimators entails
uncontrolled approximations. In particular, although
there is no rigorous way to establish the range where the leading correction
in $\epsilon$ is sufficient, or the number of powers in $\epsilon$ needed
over an arbitrary range, PW can be accurately extrapolated to the unbiased result.
Here, the bias of the PW estimator has a leading contribution of $\epsilon^2$
because $\Psi_0$ does not have odd parity across the node.
The second--order bias can be removed with a different choice 
of the polynomial, e.g. $f_\epsilon(x)=60x^2-200x^3+225x^4-84x^5$.
Here, we stick with the original estimator of  ref.~\citenum{pw}\,,
but we note that there is a freedom in the choice of $f_\epsilon$ that
can be tailored for optimal performance in specific situations.
The right panel shows that the relative efficiency may depend on the 
system at hand; in this example, it varies over the range of $a$ 
considered.

\begin{figure}[h!]
\includegraphics[width=8cm]{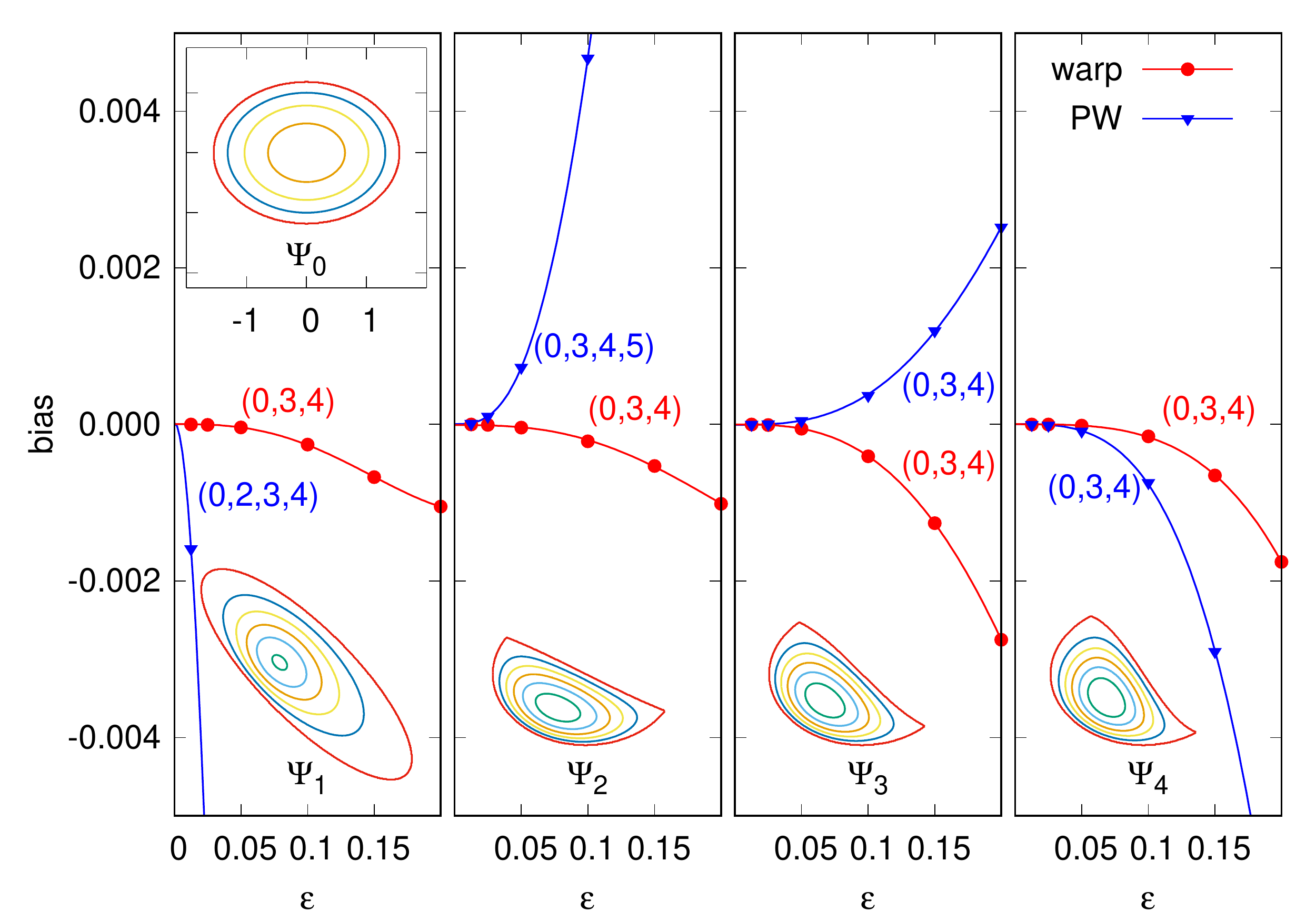}
\caption{Bias of the VMC derivative for PW and warp with approximate Hessian, 
calculated by quadrature for various wave functions $\Psi_1$--$\Psi_4$. In each 
case, we show a contour plot of the normalized $\Psi_i$ with level lines from 0 
in steps of 0.2 (the contour plot of $\Psi_0$ in the top left inset defines the
$(x,y)$ scale). The colored labels near each curve indicate the powers in
$\epsilon$ needed to extrapolate to the unbiased value with five--digit 
accuracy. 
The bias of PW for $\Psi_2$--$\Psi_4$ has a leading term $\epsilon^3$ because 
of the odd parity of the wave function across the relevant node. 
Empirically we see that 
the warp estimator has a cubic leading term in all cases, including $\Psi_1$.
}
\label{fig:sinx}
\end{figure}
The bias of the warp estimator when the off--diagonal elements of the Hessian 
are neglected is compared to the bias of the PW estimator in 
Figure~\ref{fig:sinx}\,. Since we need a non--diagonal Hessian to start with, 
we consider {\it (i)} a rotated, more eccentric ellipse with a further 
non--symmetrical distortion of the wave function, and {\it (ii--iv)} 
the positive lobe of a wave function limited to half of the original 
ellipse by the nodal line $y+\sin(\alpha x)=0$, with $\alpha=0.5$, 1 and 1.5
(see the contour plots in Figure~\ref{fig:sinx}\,). 
The derivative is taken with respect to $a$ for $\Psi_1$,
and with respect to $\alpha$ for $\Psi_2$--$\Psi_4$.
Within this (very limited) set of test cases, the bias is smaller and less
system--dependent for the warp than for the PW estimator. 
The important result is that neither involves uncontrolled approximations,
as both can be extrapolated to the unbiased result in a single run.
We have also verified that the warp 
estimator with the full Hessian is unbiased for finite $\epsilon$.

\subsection{Diffusion Monte Carlo}
We now consider the derivative in DMC. The FN--DMC algorithm is a branching
random walk of many weighted walkers, generated by a short--time approximation
$G(R',R)$ to the importance--sampled Green's function, which asymptotically 
samples the distribution $P(R)=\Psi(R)\Phi(R)$.\,\cite{reviews}
The problem with the derivative estimator, eq~\ref{eq:de_bare}\,,
is the presence of the logarithmic 
derivative of $P(R)$, which is not a known function of $R$. However, 
$P$ is the marginal distribution of the joint probability density $P_{\rm joint}$
of the whole random walk, which does have an explicit expression as
a product of Green's functions,
\begin{eqnarray}
&P(R_n)=\int dR_0\ldots dR_{n-1}P_{\rm joint}(R_n,R_{n-1}\ldots,R_0)&\nonumber\\
&\equiv \int dR_0\ldots dR_{n-1} \prod_{i=0}^{n-1}G(R_{i+1},R_i)P_0(R_0)& 
\label{eq:marginalization}
\end{eqnarray}
where $P_0$ is the (largely arbitrary) probability distribution of 
the initial configuration $R_0$. Therefore, it is sufficient to consider
the estimator of $d_\lambda E$ in eq~\ref{eq:de_bare}\,
as an average over the whole trajectory of the random walk, rather than over
the current configuration, to bring an explicitly known probability
distribution to the fore.\,\cite{vd} This is similar to the calculation of
forces in path integral Monte Carlo.\,\cite{zong}

In practice, the inclusion of the entire trajectory in the estimator 
is not necessary.
As shown in ref.~\citenum{vd}\,.
the logarithmic derivative of the DMC density distribution $P$ in the 
estimator of eq~\ref{eq:de_bare}\, can be 
replaced by the summation
\begin{equation}
d_\lambda\ln P(R_n)=d_\lambda\sum_{i=n-k}^{n-1}\ln G(R_i',R_i)
\label{eq:pimc}
\end{equation}
over the last $k$ steps of the random walk, with $R_n$ the current 
configuration $R$ of eq~\ref{eq:de_bare}\,.
The omitted term,\,\cite{vd} 
$\big\langle\left[E_L(R_n)-E\right]d_\lambda\ln P(R_{n-k})\big\rangle$,
vanishes for sufficiently large $k$ because $E_L(R_n)$ and $d_\lambda\ln P(R_{n-k})$
become statistically independent variables and $\big\langle E_L-E\big\rangle=0$.

In eq~\ref{eq:pimc}\,, $G(R_i',R_i)$ is the transition rule from $R_i$ 
to $R_i'$ of the random walk. It includes a Metropolis
test to reduce the time step error;\,\cite{reviews} 
therefore, $R_i'$ is the 
configuration proposed when the walker is at $R_i$, and the
next configuration $R_{i+1}$ is $R_i'$ or $R_i$ if the move is accepted
or rejected, respectively. 
Note that in the formal expression of $P_{\rm joint}$, eq~\ref{eq:marginalization}\,,
the arguments of the Green's functions are integrated over, whereas in the
contribution to the estimator of the derivative, eq~\ref{eq:pimc}\,, they
are the particular values of the particles' coordinates effectively sampled 
by the random walk. Correspondingly, the actual value taken by $G(R',R)$ is
\begin{widetext}
\begin{equation}
\begin{cases}
T(R',R)p(R',R)W(R',R) &\text{for an accepted move}\\
T(R',R)[1-p(R',R)]W(R,R)  &\text{for a rejected move} 
\end{cases}
\label{eq:g}
\end{equation}
\end{widetext}
where $T$ is the a--priori transition probability, $p$ the probability
of accepting the move, and $W$ the branching factor (see below).

The inclusion of rejected configurations in eq~\ref{eq:pimc}\, and of
the factor $p$ or $1-p$ in eq~\ref{eq:g}\, in the derivative of the
full Green's function are instrumental to obtain an estimate of
$d_\lambda E$ completely consistent with the DMC energy $E(\lambda)$ calculated 
at the same time step. Their omission still can give an unbiased result
in the limit $\tau\to 0$, but it may cause an unacceptably large time step 
error on the derivative.\,\cite{vd}

The functions $T$, $p$, and $W$ are standard:\,\cite{umrigar93}
\begin{widetext}
\begin{eqnarray}
T(R',R)&=&\exp\left\{-\left[R'-R-F(R)V(R)\tau\right]^2/2\tau\right\}\nonumber\\
p(R',R)      &=&
\begin{cases}
0 &\text{if the move crosses a node}\\
\min\left\{1,\left[\Psi^2(R')T(R,R')\right]\Big/\left[\Psi^2(R)T(R',R)\right]
\right\}&\text{otherwise}
\end{cases} \\
W(R',R)&=&\exp\left\{[S(R')+S(R)]\tau/2\right\}.\nonumber
\end{eqnarray}
\end{widetext}
Here, $\tau$ is the time step, $V=\nabla\ln\Psi$ is the so--called velocity,
and $F(R)=\sqrt{2V^2\tau-1}/(V^2\tau)$ is the damping factor of its divergence
near the nodes; 
the logarithm of the branching factor is also damped at the nodes,
$S(R)=\left[E_{\rm est}-E_L(R)\right]F(R)-\ln(N/N_0)$, where
$E_{\rm est}$ is the best current estimate of the energy and $N$ and
$N_0$ are the current and the target number of walkers. 

The presence of $E_{\rm est}$, the Monte Carlo estimate of $E$, in the 
branching term $S$ implies that the calculation of $d_\lambda E$ includes a 
contribution proportional to $d_\lambda E$ itself:
\begin{equation}
\tau \Big\langle\sum_i [E_L(R_n)-E]F(R_{n-i})\Big\rangle d_\lambda E\equiv 
\overline{F}d_\lambda E 
\label{eq:deest}
\end{equation}
This does not require prior knowledge of the result: we can calculate the 
factor $\overline{F}$ and $(d_\lambda E)_0$, the derivative when the contribution of 
eq~\ref{eq:deest}\, is omitted, and combine them to get the unbiased result 
as $d_\lambda E=(d_\lambda E)_0+\overline{F}d_\lambda E$, or $d_\lambda E=(d_\lambda E)_0/(1-\overline{F})$.

The main technical complication in DMC is the need to store
a few quantities for each derivative and for each value 
of $\epsilon$ 
over the last $k$ steps of each walker, namely
$d_\lambda \ln G$ and $E_L d_\lambda \ln G$ to implement eq~\ref{eq:de_bare}\, 
with the probability distribution $P$ of eq~\ref{eq:pimc}\,, and
$F$ and $E_LF$ to implement eq~\ref{eq:deest}\,.

The AS regularized estimator has been applied to approximate DMC forces
in refs.~\citenum{vd, valsson}\,. However it pushes a finite 
density of walkers on the nodes, which is presumably not optimal in DMC. 
Furthermore, unlike in VMC, it requires\,\cite{vd} an extrapolation to 
$\epsilon\to 0$ which --at difference with the PW and warp estimators-- 
cannot be done in a single run.

Therefore, for the DMC derivatives we consider only the PW and the warp 
estimators. For the former, we insert eq~\ref{eq:pimc}\, into
eq~\ref{eq:de_bare}\,, and multiply each term of the resulting summation
by a polynomial $f_\epsilon$ calculated at the appropriate configuration.
For the latter, we 
just insert
eq~\ref{eq:pimc}\, in eq~\ref{eq:de_warp}\,. 
For the warp estimator, the argument of the Jacobian needs some care: 
we evaluate $\partial_\lambda \ln J$ in the proposed configuration 
$R'$ for both accepted and rejected moves; alternatively,
we can include $\partial_\lambda \ln J$ only for the accepted moves,
provided 
the warp transformation is not considered in the
derivatives
at $R'$ when the move is rejected.

We present results of DMC simulations with the wave function $\Psi_0$,
time step $\tau=0.1$, and target number of walkers $N_0=100$. 
We have verified that in the limit $\tau\to 0$ we recover the
analytic results\,\cite{book} for the ground state energy,
$E=2q/a^2$ with $q=0.825352549$, within a statistical 
error of less than one part in 10,000. To this level of accuracy, 
the population control bias\,\cite{reviews} is negligible.

\begin{figure}[h!]
\includegraphics[width=8cm]{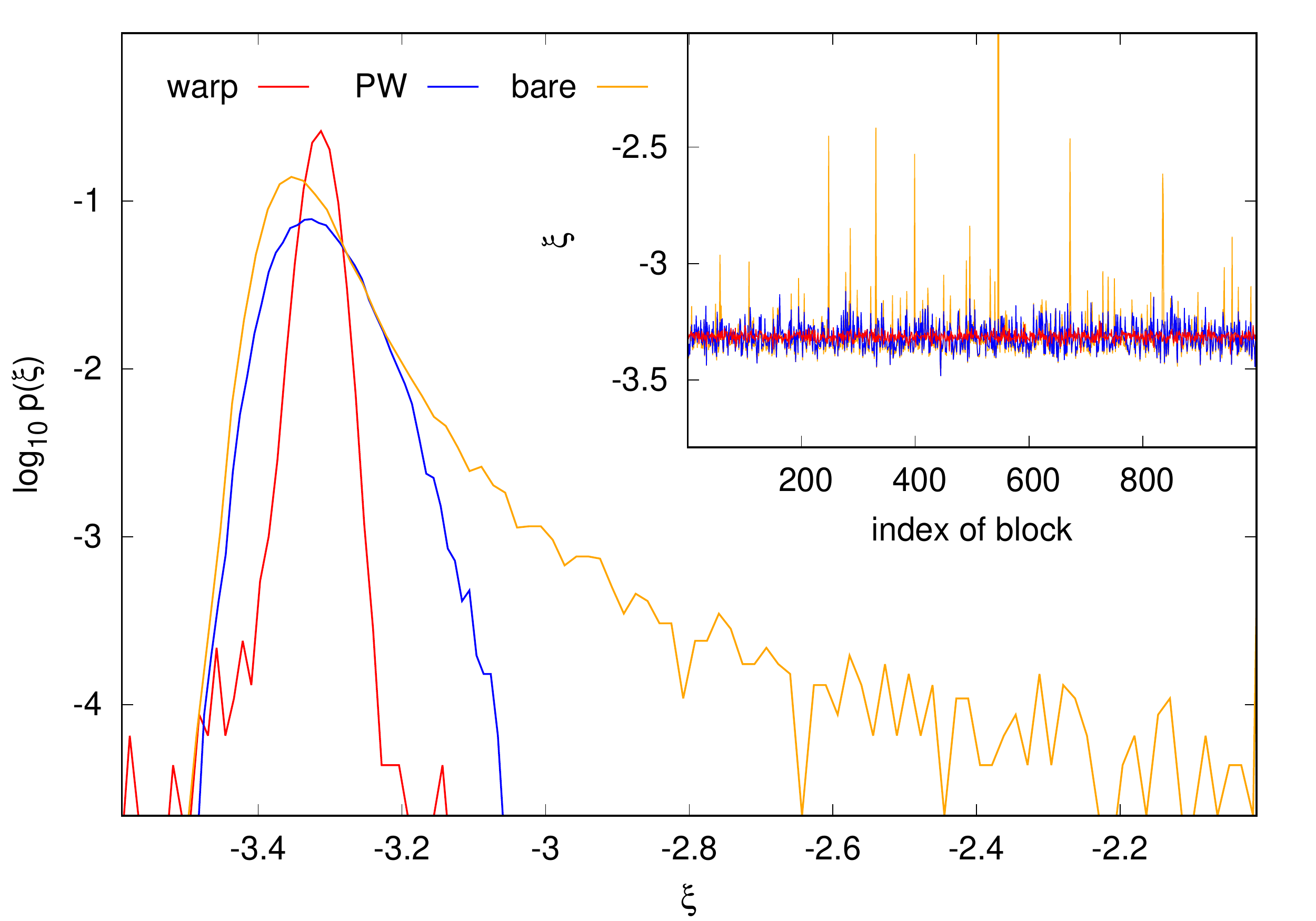}
\caption{Histogram of 46,000 block averages, each of which taken over 10,000 
steps of 100 walkers in a DMC calculation of $d_aE\big|_{a=1}$. 
The warp and PW estimators (with $\epsilon=0.2$ and 0.0125, respectively)
have nearly Gaussian distributions. The bare estimator has a heavy--tailed 
distribution; the largest value in the present sample exceeds $x=10$.
Inset: the data trace of the first 1,000 block averages.
\label{fig:histo}
}
\end{figure}
Figure~\ref{fig:histo}\, exposes the drawback of the bare estimator:
the probability distribution $p(\xi)$ for the block averages of the 
derivative features a right heavy tail, consistent with the 
expected\,\cite{trail2,badinski} leading decay $\propto\big|\xi-\xi_0\big|^{-5/2}$.
In the data trace, shown in the inset, heavy tails result in large spikes
that would mar the smooth convergence of structural or variational optimization.
Meaningful averages and statistical uncertainties of heavy--tailed 
distributions with known tail indices can be computed with a tail regression 
analysis.\,\cite{tre} 
This technique, however, requires a heavy post--processing not very practical
for large--scale applications. The regularized estimators PW and warp, instead, 
have nearly Gaussian distributions amenable to standard statistical analysis
with significantly smaller statistical errors and, most importantly, no
large spikes in the data trace.

\begin{figure}[h!]
\includegraphics[width=8cm]{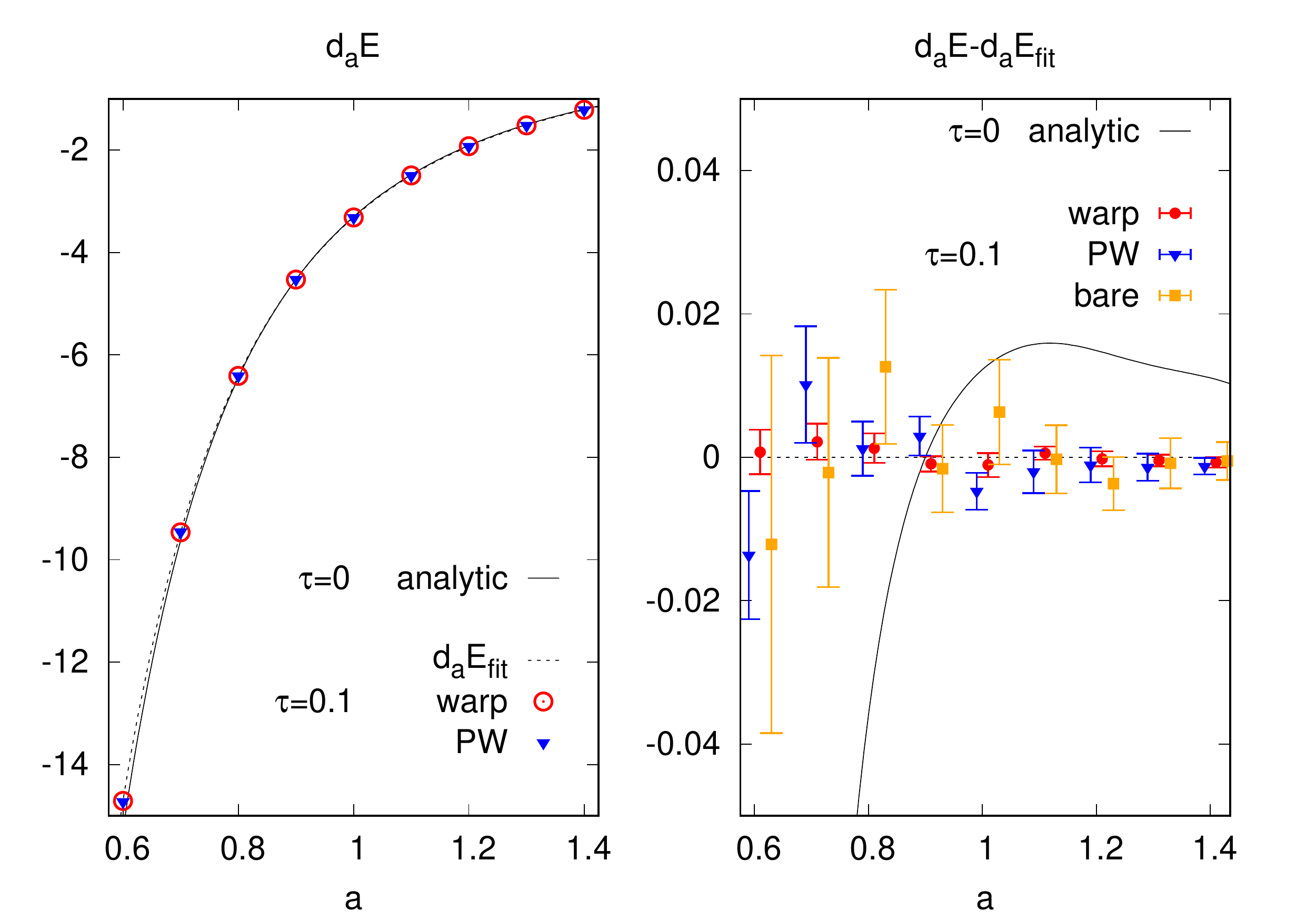}
\caption{DMC calculations of $d_aE$ with $\tau=0.1$.
Left panel: the warp and PW results compare favorably with the
``exact'' result, defined as the derivative $d_aE_{\rm fit}$ of a fit to DMC 
calculations of $E(a)$. The analytic result $d_aE=-4q/a^3$
differs from $d_aE_{\rm fit}$ because the latter has a finite time step
error.
Right panel: the difference of the calculated derivatives with $d_aE_{\rm fit}$
are shown on an expanded scale by subtracting the latter
(hence the solid black line is minus the time step error of 
$d_aE_{\rm fit}$).
The PW derivatives are extrapolated to $\epsilon\to 0$.
We also include the bare--estimator derivatives, with averages and
statistical uncertainties obtained with the tail regression estimator 
analysis toolkit made available in ref.~\citenum{tre}\,.
Small horizontal shifts are applied to same--$a$ data for clarity.
}
\label{fig:dmc}
\end{figure}
The central result of this work is shown in Figure~\ref{fig:dmc}\,. We
calculate the energy $E$ and its derivative $d_aE$ for a set of values of 
$a$, and compare the DMC derivatives with the derivative of a fit to the
DMC energies. All the estimators (bare, extrapolated PW, and warp)
are unbiased, which demonstrates the correctness of the proposed algorithm.
For comparison, the variational drift--diffusion (VD) approximation 
of ref.~\citenum{vd}\, gives for
$a=1$ a bias of $\sim 0.2$, twice the full scale of the right panel, and it
gets even worse for smaller time steps (although the VD 
approximation is devised to exploit good wave functions,
while our $\Psi_0$ is poor on purpose to test the unbiased estimators).

For given $a$, the same run is used for all the estimators. Therefore
the statistical error is a direct measure of the square root of their 
relative efficiency. The statistical errors of the bare, PW, and warp 
estimators, averaged over the values of $a$ shown in Figure~\ref{fig:dmc}\,, 
are in the ratio 4.9:2.4:1. These figures may belittle the PW estimator
somewhat, because in this particular example a large quadratic bias
needs to be eliminated by extrapolation, but they convey the relevant message 
that both PW and warp are significantly more efficient than the bare estimator.

Finally, the ratio between the statistical error of the DMC and VMC
derivatives calculated with the warp estimator at $\epsilon=0.2$, using
the same time step and the same number of Monte Carlo samples,
lies between 1.9 and 1.7 in the range of $a$ of Figure~\ref{fig:dmc}\,.
We consider this ratio a favorable indication of the efficiency
of the algorithm, which will hopefully spur a full assessment 
with realistic many--body wave functions.

\section{Conclusions}
In summary, we have presented an algorithm to calculate unbiased, 
finite--variance derivatives in DMC. The estimate of the derivative with
respect to a given parameter is fully consistent with the dependence
on that parameter of the FN energy, calculated with the same time step. 
The tail regression statistical analysis\,\cite{tre} can cope with
the problem of the infinite variance of the bare estimator. Alternatively,
and more efficiently, both the recently proposed PW regularization\,\cite{pw}
and the warp regularization introduced in this work can be used to good 
effect to eliminate the divergence of the variance.
 
\section*{Acknowledgments}
CF and SM acknowledge support from the European Centre of Excellence 
in Exascale Computing TREX, funded by the European Union's Horizon 2020 - Research and
Innovation program - under grant no. 952165.

\bibliography{w.3}

\end{document}